\documentclass{appolb}
\usepackage{epsfig}
\usepackage{amssymb}
\usepackage{amsmath}
\usepackage{cite}

\newcommand{\eps}{\varepsilon}
\newcommand{\gsim}{\raisebox{-0.07cm   }
{$\, \stackrel{>}{{\scriptstyle\sim}}\, $}}

\begin{document}
\title{{\small \rm DESY 08--067,  \hfill }\\
Higher order corrections to heavy flavour production in 
deep inelastic scattering
\thanks{Presented at Cracow Epiphany Conference 2008}%
}
\author{I. Bierenbaum, J. Bl\"umlein and S. Klein 
%
\thanks{This paper was supported in part by SFB-TR-9: Computergest\"utze 
Theoretische Teilchenphysik, and the Studienstiftung des Deutschen 
Volkes.}
%
\vspace{.3cm}\\
%
Deutsches Elektronen-Synchrotron, DESY,
Platanenallee 6, D-15738 Zeuthen, Germany
}

\maketitle
\begin{abstract}
  In the asymptotic limit $Q^2 \gg m^2$, the non-power corrections to
  the heavy flavour Wilson coefficients in deep--inelastic scattering
  are given in terms of massless Wilson coeffcients and massive
  operator matrix elements. We start extending the existing NLO
  calculation for these operator matrix elements by calculating the
  O($\varepsilon$) terms of the two--loop expressions and having first
  investigations into the three--loop diagrams needed to
  O($\alpha_s^3$).
\end{abstract}
\PACS{PACS numbers come here}
  
\section{Introduction}
In deep--inelastic scattering, the differential cross-section with
respect to the Bjorken--variable x and the virtuality of the photon
$Q^2$, can be expressed in terms of the unpolarized structure
functions $F_2(x,Q^2)$ and $F_L(x,Q^2)$, and the polarized structure
functions $g_1(x,Q^2)$, $g_2(x,Q^2)$. For small values of x, the
contributions of heavy charm product to $F_2(x,Q^2)$,
$F_2^{c\bar{c}}(x,Q^2)$, are of the order of 20 - 40 \%, and therefore
deserve and need a more detailed investigation. So far there exist NLO
-- 2--loop -- heavy flavour corrections to $F_2^{p,d}(x,Q^2)$ in the
whole kinematic range, calculated in a semi--analytic way in
$x$--space \cite{HEAV1}.  A fast implementation for complex Mellin
$N$--space was given in \cite{HEAV2}.  One observes that
$F_2^{c\bar{c}}(x,Q^2)$ is very well described by an asymptotic result
for $F_2^{c\bar{c}}(x,Q^2)|_{Q^2\gg m^2}$ for $Q^2 \gsim 10 m_c^2$.
For these higher values of $Q^2$, one can calculate the heavy flavour
Wilson coefficients, the perturbative part of the structure functions
$F_2(x,Q^2)$ and $F_L(x,Q^2)$, analytically, which has been done for
$F_2(x,Q^2)$ to 2--loop order in \cite{BUZA,BBK1} and for $F_L(x,Q^2)$
to 3--loop order in \cite{BFNK}. First steps towards an asymptotic
3--loop calculation for $F_2^{c\bar{c}}(x,Q^2)$ have been done by the
authors by calculating the first O($\eps$) terms of the 2--loop
diagrams \cite{UnPolOeps}, contributing to 3--loop heavy flavour
Wilson coefficients via renormalization. We report here on further
steps towards a full 3--loop calculation.

\section{Heavy flavour Wilson coefficients in the limit $Q^2 \gg m^2$}

On the twist--2 level, the structure functions can be expressed as a
convolution of perturbatively calculable Wilson coefficients and
non--perturbative parton densities. We consider here the heavy flavour
contributions to these Wilson coefficients, the heavy flavour Wilson
coefficients.
In the region $Q^2 \gg m^2$, one can use the massive renormalization group
equation to obtain all non--power corrections to these heavy flavour Wilson
coefficients as convolutions of massless -- light -- Wilson coefficients
$C_k(Q^2/\mu^2)$ and massive operator matrix elements (OMEs)
$A_{ij}(\mu^2/m^2)$ \cite{HEAV1}.
The light Wilson coefficients are known by now up to three loops \cite{MVV}
and carry all the process dependence. The operator matrix elements, on the
other hand, are universal, process--independent objects, which are calculated
as flavour decomposed operators in light--cone expansion between partonic
states. Both objects have an expansion in $\alpha_s$, where one should note
that the external legs of the diagrams contributiong to the OMEs are on-shell.

\section{Massive operator matrix elements}

In order to perform the 3--loop calculation of the OMEs, one has to
first calculate the bare quantities and then to renormalize them,
where they need to be mass-- and charge--renormalized and contain
ultraviolet (UV) and collinear divergences.  The mass renormalization
is done in the on--shell scheme \cite{MASS1,MASS2}, whereas the charge
renormalization is done using the $\overline{\rm{MS}}$ scheme. After
mass-- and charge--renormalization, the remaining UV--divergences are
accounted for by operator renormalization via Z--factors, $Z_{ij}$,
and the collinear divergences via mass factorization, multiplying by
$\Gamma_{ij}$. The Z--factors are given by the generic formula:\\
\includegraphics[width=12.5cm]{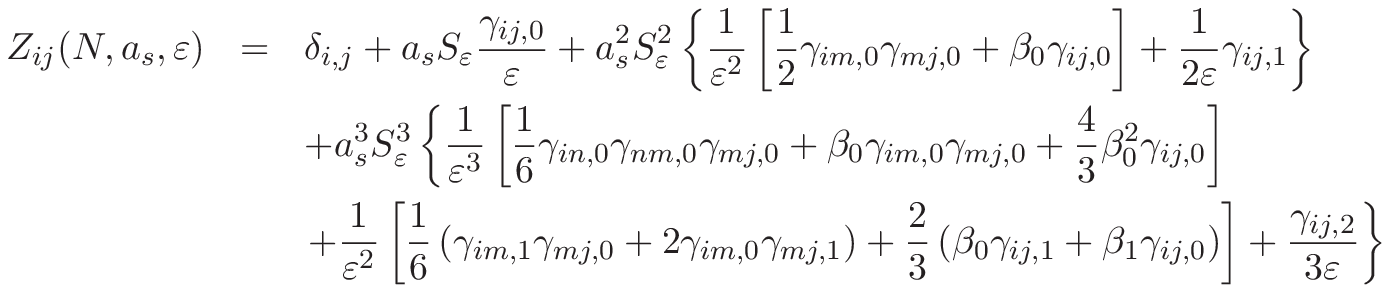}\\ which has to be
adapted for the various flavour decomposed combinations. The indices
$i,j$ here either run over $i,j \in \{q,g\}$ or denote the
non--singlet combination. The pure--singlet Z--factor is given by:
$Z_{qq}^{PS}=Z_{qq}-Z_{NS}$. $\gamma_{ij,k}$ are the k--loop anomalous
dimensions and $\beta_i$ denote the expansion coefficients of the beta
function.  The $\Gamma_{ij}$ are in the massless case the inverses of
the Z--factors, however, since we always have at least one heavy quark
loop, they differ and the $\Gamma_{i,j}$ contribute at most to
two--loop subdiagrams.\\
Let us consider, e.g., the renormalized gluonic matrix element
$A_{Qg}$, where $\hat{A}_{Qg}$ denotes the mass-- and
charge--renormalized expression. One finds then:
\\[1em]
\includegraphics[width=1.25cm,angle=90]{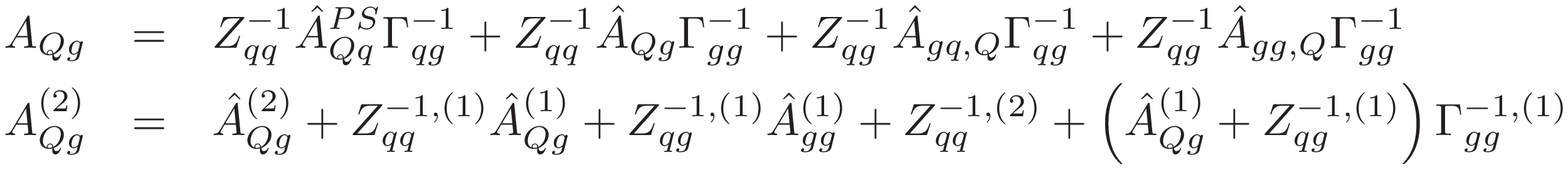}\\[1em] As the
general expression in the first line already indicates, there is a
mixing with {$\hat{A}_{Qq}^{PS}$} and {$\hat{A}_{gg,Q}$} from
O($\alpha_s^3$), while {$\hat{A}_{gq,Q}$} starts contributing from
O($\alpha_s^2$) and therefore at O($\alpha_s^4$) to
{$\hat{A}_{Qg}$}. For ${A}^{(2)}_{Qg}$ given in the second line, one
finds the $O(\eps)$ term of the gluonic one--loop OME,
${\overline{a}_{Qg}^{(1)}}$, entering the two--loop expression via
renormalization, as described for example in
\cite{BUZA,BBK1,UnPolOeps}. In as the same way, for the
renormalization of ${ A_{Qg}^{(3)}}$, the O($\eps$)--term of the
two--loop expression ${\overline{a}_{Qg}^{(2)}}$ is needed -- as are
the terms ${\overline{a}_{gg}^{(2)}}$, ${\overline{a}_{Qq}^{(2),PS}}$
due to the above mentioned operator mixing, and the O($\eps^2$) of the
one--loop ${A_{Qg}^{(1)}}$. The calculation of these O($\eps$) terms
is a first step towards a 3--loop calculation, cf. section 3.1.,
while first calculations of 3--loop diagrams for fixed Mellin N are
described in section 3.2.\\ As a last remark, note that we consider
charm quark contributions here, while for heavier quarks decoupling
\cite{DECOUP} has to be applied.

\subsection{Two--loop diagrams to O($\eps$) for general Mellin N}

Our calculation is performed in Mellin space, where the convolution of
functions becomes a simple product. The O($\eps$) terms for the
unpolarized gluonic OMEs, as for the pure--singlet and non--singlet
cases, have been given in \cite{UnPolOeps} for general Mellin N. The
corresponding polarized contributions are to be published soon. The
calculation is performed in two ways: on the one hand, we rewrote the
OMEs in terms of Mellin--Barnes integrals and used M. Czakon's
mathematica package MB \cite{MB} to obtain numeric results, serving as
a check for the analytic results, which have been obtained expressing
the OMEs as generalized hypergeometric functions. Expanding these
functions in $\eps$, one has to re-sum the expression for the desired
order, which we did using integral techniques and C. Schneider's
mathematica package SIGMA \cite{sigma}. The results are then given in
terms of nested harmonic sums \cite{HS1,HS2}, to which we applied
algebraic and analytic simplifications \cite{ALGEBRA,STRUCT} to find
the easiest possible representation.\\ The term $A_{gg}^{(3)}$ has
been newly calculated and
is given up to order O($\eps$) by:\\[1em]
\includegraphics[width=6.5cm,angle=90]{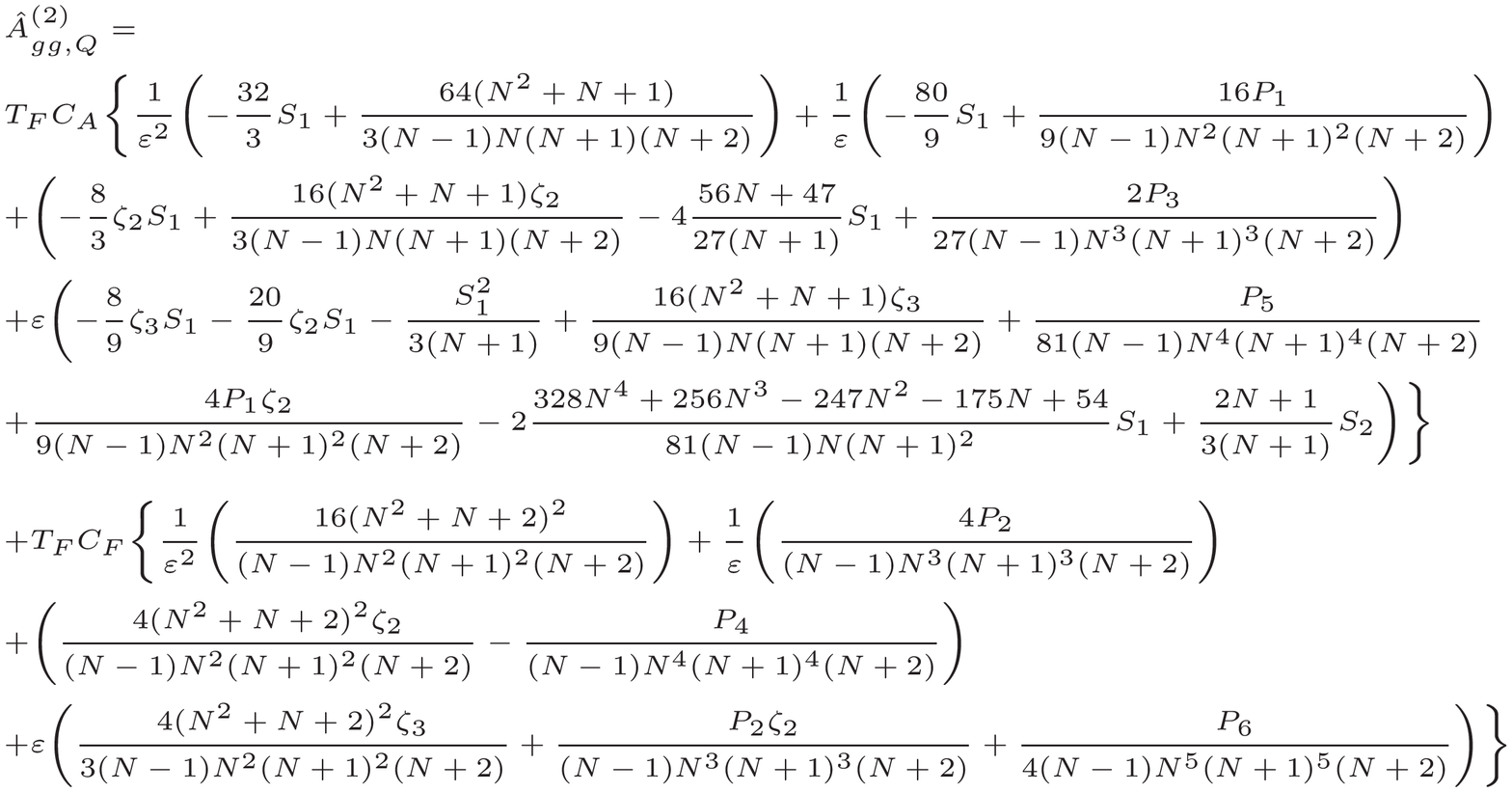}\\[1em]
\includegraphics[width=12.4cm]{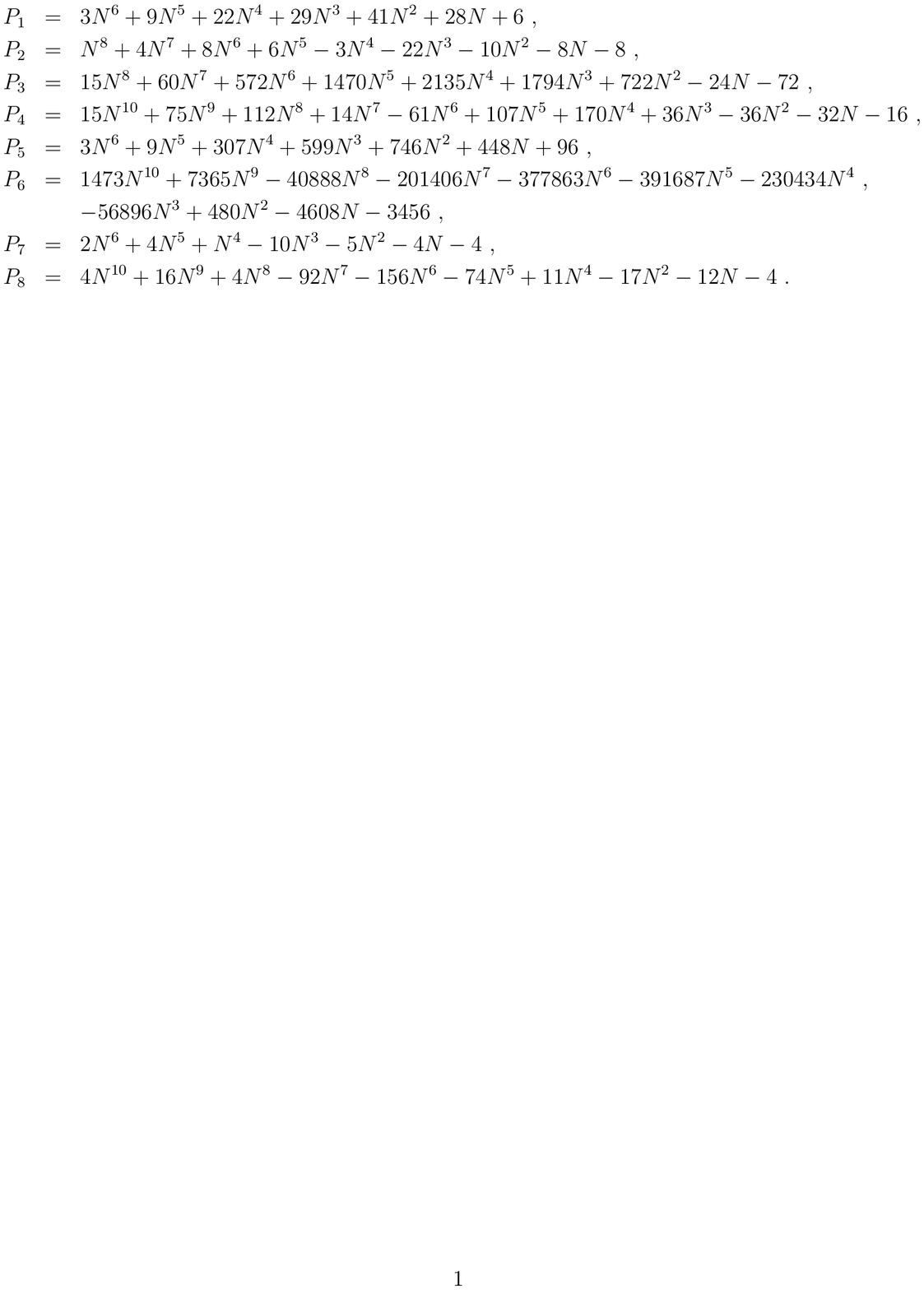}\\ [0.5em] where we have
calculated an all order $\eps$ result, which is solely given in terms
of Euler $\Gamma$ and $\psi$ functions. This expression is also needed
in the context of the variable flavour number scheme [].  \\ In the
unpolarized case, all 2--loop O($\eps$) terms are now known. In the
polarized case, the calculation proceeds in the same way and we
calculated so far the gluonic, pure--singlet and non--singlet terms,
which will be published soon \cite{UnPolOeps}.

\subsection{Fixed values of N at three loops}

As a next step towards a full O($\alpha_s^3$) calculation, we started
calculating unpolarized three--loop OMEs $A^{(3)}_{ij,Q}$ for fixed
values of Mellin N. The contributing OMEs are: singlet: \{$A_{Qg}$,
$A_{gg,Q}$, $A_{gq,Q}$\}, pure-singlet: $A^{PS}_{Qq}$, non-singlet:
\{$A^{NS,+}_{qq,Q}$, $A^{NS,-}_{qq,Q}$, $A^{NS,v}_{qq,Q}$\}, where we
have operator mixing between the singlet and pure-singlet terms. The
first object of investigation is the gluonic $A^{(3)}_{Qg}$: The
necessary three--loop diagrams are generated using QGRAF
\cite{Nogueira}, where the operator product expansion has been
implemented up to insertions of operators with three and four gluonic
lines. The number of diagrams contributing to $A^{(3)}_{Qg}$, e.g., is
1478 diagrams with one and 489 diagrams with two quark loops, where at
least one of the loops is heavy.  \\ The steps for the calculation of
these self-energy type diagrams with one additional scale set by the
Mellin variable N, are the following: The diagrams are genuinely given
as tensor integrals due to the operators contracted with the
light--cone vector $\Delta$, $\Delta^2=0$.  The idea is, to first undo
this contraction and to developed a projector, which, applied to the
tensor integrals, provides the results for the diagrams for a specific
(even) Mellin N under consideration. So far, we implemented the
projector for the first 4 contributing Mellin N, $\rm{N}=2,...,8$,
where the color factors are calculated using \cite{COLORF}. The
diagrams are then translated into a form, which is suitable for the
program MATAD \cite{MATAD} by M.~Steinhauser, which does the expansion
in $\eps$ for the remaining massless and massive three--loop
tadpole--type diagrams. We have implemented these steps into a FORM
\cite{FORM} program and tested it against some of our two--loop
results and the all-order $\eps$ result of $A^{(2)}_{gg,Q}$ and found
agreement. We then turned to a subset of the 3--loop diagrams, the
diagrams $\propto T_F^2$: $T_F^2C_F$, $T_F^2C_A$, which are currently
under investigation. 

\section{Conclusions and outlook}

We calculated the O($\eps$) contributions to heavy flavour Wilson
coefficients for general Mellin variable N at O($\alpha_s^2$), as a
first step towards a O($\alpha_s^3$) calculation. Furthermore, we
installed a program chain to calculate the corresponding 3--loop
diagrams to O($\alpha_s^3$), with the help of MATAD. This chain is now
existing and we expect first results in the near future.



\end{document}